\documentclass[traditabstract]{aa}  
\usepackage{graphicx}
\usepackage{txfonts}
\begin{document}
   \title{IGR J12319--0749: evidence for another extreme blazar found with \emph{INTEGRAL}}

%   \subtitle{IGR J12319--0749: another extreme blazar found with \emph{INTEGRAL}}

    \titlerunning{IGR J12319--0749: evidence for another extreme blazar found with INTEGRAL}
\authorrunning{L.~Bassani}

   \author{L. Bassani\inst{1}, R.~Landi\inst{1}, F. E.~Marshall\inst{2}, A. Malizia\inst{1}, 
A. Bazzano\inst{3},
A. J. Bird\inst{4}, N. Gehrels\inst{5}, P. Ubertini\inst{3}, N. Masetti\inst{1}}
   \offprints{bassani@iasfbo.inaf.it}
\institute{INAF/IASF Bologna, via Piero Gobetti 101, I-40129 Bologna, Italy 
\and
Astrophysics Science Division,
NASA Goddard Space Flight Center, Greenbelt, MD 20771, USA \and
INAF/IAPS Rome, Via Fosso del Cavaliere 100, I-00133 Rome, Italy \and
School of Physics and Astronomy, University of Southampton, Highfield, SO17 1BJ, UK}

   \date{Received  / accepted}
% \abstract{}{}{}{}{} 
% 5 {} token are mandatory

\abstract
{We report on the identification of a new soft gamma-ray source, IGR 
J12319--0749, detected with the IBIS imager on board the \emph{INTEGRAL} satellite. The source, which 
has an observed 20--100 keV flux of $\sim$8.3 $\times$ 10$^{-12}$ erg cm$^{-2}$ s$^{-1}$, is 
spatially coincident with an AGN at redshift $z=3.12$. The broad-band continuum, obtained by combining 
XRT and IBIS data, is flat ($\Gamma$=1.3) with evidence for a spectral break around 25 keV (100 keV in 
the source rest frame). X-ray observations indicate flux variability which is further supported by a 
comparison with a previous ROSAT measurement. IGR J12319--0749 is also a radio emitting object likely 
characterized by a flat spectrum and high radio loudness; optically it is a broad-line emitting object 
with a massive black hole ($2.8\times10^{9}$ solar masses) at its center. The source Spectral 
Energy Distribution is similar to another high redshift blazar, 225155+2217 at $z=3.668$:
both objects are bright, with a large accretion disk luminosity 
and a Compton peak located in the hard X-ray/soft gamma-ray band. 
IGR J12319--0749 is likely the second most distant blazar detected so far by \emph{INTEGRAL}.}

\keywords{gamma-rays: observations, X-rays: galaxies - galaxies: active: individual: IGR J12319--0749
}

\maketitle 

\section{Introduction}
Blazars are the most powerful of all Active Galactic Nuclei (AGNs); their 
continuously emitting radiation covers the entire electromagnetic spectrum from radio to gamma-ray 
frequencies. Because of their enormous luminosities, blazars are visible to very large 
distances/redshifts. In the widely adopted scenario of blazars, a single population of high-energy 
electrons in a relativistic jet radiate from the radio/FIR to the UV/soft-X-ray through the synchrotron 
process and at higher frequencies through inverse Compton (IC) scattering soft-target photons present either 
in the jet (synchrotron self-Compton [SSC] model), in the surrounding material (external Compton [EC] 
model), or in both (Ghisellini et al. 1998 and references therein). Therefore a strong signature of the 
blazar nature of a source is a double peaked structure in the Spectral Energy Distribution (SED), with 
the synchrotron component peaking anywhere from infrared to X-rays and the Compton component extending 
up to GeV or even 
TeV gamma-rays. To explain all the different SEDs observed in blazars, Fossati et al. (1998) proposed the 
blazar sequence, which claims an inverse relation between peak energies and source luminosity: the more 
luminous sources have both synchrotron and Compton peaks at lower energies than their fainter (and 
generally 
at lower redshifts) counterparts. Within the blazar population, high redshift objects, which belong to 
the class of flat spectrum radio Quasars (FSRQ), tend to be the most luminous ones. Recently, Ghisellini 
et al. (2010) showed that the hard X-ray selected blazars at high 
redshifts  are those with the  most powerful jets, the most luminous accretion disks and 
the largest black hole masses. In other words, they are among the most extreme blazars, even more extreme 
than those selected in other wavebands like the gamma-ray region explored by \emph{Fermi}/LAT.

Theoretically, this can be understood on the basis of the blazar sequence and indeed can be considered a 
proof of its validity: the high energy peak in the SED of the most powerful blazars is in the hard 
X-ray/MeV range. As a consequence, these objects are more luminous in the 20--100 keV band than above 100 
MeV, and thus become detectable in the hard X-ray surveys even if they are undetected in the gamma-ray 
ones. Taken from another point of view, this also means that a survey in hard X-rays is more efficient in 
finding the most powerful blazars lying at the highest redshifts.

Within the \emph{Swift}/BAT sample, there are 10 blazars (all FSRQs) at redshift greater than 2, of which 
five have a redshift between 3 and 4 (Ghisellini et al. 2010). Within the 
\emph{INTEGRAL}/IBIS surveys we have 8 blazars at $z\ge 2$ of which 2 are located between 3 and 4 
(Malizia et al. 2012).

Here, we provide evidence for the discovery with \emph{INTEGRAL} of yet another powerful blazar, 
IGR J12319--0749 associated with a quasar  at $z=3.12$; we present a set of follow-up observations 
with \emph{Swift}/XRT/UVOT, provide the first combined X/gamma-ray spectrum of the source and discuss its 
non simultaneous SED. This is likely the second most distant blazar ever detected by 
\emph{INTEGRAL}.

\section{\emph{INTEGRAL} data}
IGR J12319--0749 was reported for the first time as a high energy emitter by Bird et al. (2010) in 
the 4th \emph{INTEGRAL}/IBIS catalogue. The source is detected with a significance of $\sim$4.8$\sigma$
at a position corresponding to R.A.(J2000) = $12^{\rm h}31^{\rm m}54^{\rm s}.5$ 
and Dec.(J2000) = $-07^{\circ}48^{\prime}57^{\prime\prime}.6$ with an associated positional uncertainty
of $4.7^{\prime}$ (90\% confidence level); the source location is 54.8 degrees above the Galactic plane,
strongly suggesting an extragalactic nature.

The IBIS spectrum was extracted using the standard Off-line Scientific Analysis
(OSA version 7.0) software released by the Integral Scientific Data Centre.
Here and in the following, spectral
analysis was performed with XSPEC v.12.6.0  package and errors are
quoted at 90\% confidence level for one interesting parameter ($\Delta\chi^{2}=2.71$).
A simple power law provides a good fit to the IBIS data ($\chi^2$=10.4 for 8 d.o.f.)
with a photon index $\Gamma$=2.80$^{+1.35}_{-0.94}$ and an observed 20--100 keV
flux of $8.3 \times 10^{-12}$ erg cm$^{-2}$ s$^{-1}$.

\begin{figure} 
\centering 
\includegraphics[width=1.\linewidth,angle=0]{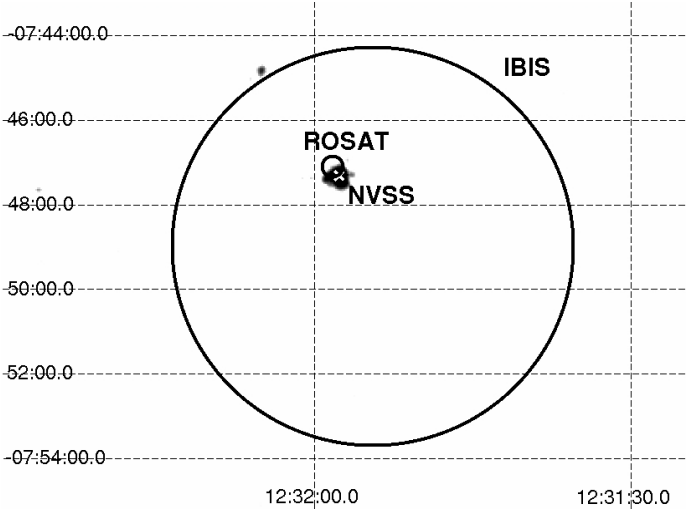} 
\caption{ XRT 0.3--10 keV 
image of the region surrounding IGR J12319--0749. The big circle corresponds to the IBIS positional 
uncertainty, while the cross indicates the location of the radio source NVSS J123157--074717; the small 
circle represents the positional uncertainty of the faint ROSAT source.} 
\label{fig1} 
\end{figure}

Within the IBIS positional uncertainty, we find a Faint ROSAT Source 1RXS J123158.3--074705, located 
with an uncertainty of $15^{\prime\prime}$ (Voges et al. 2000).
This object probably coincides with the radio source NVSS 
J123157--074717 (Condon et al. 1998), which is located $15^{\prime\prime}.8$ away, i.e. just outside 
the ROSAT error circle (see Figure~\ref{fig1}); this radio source has recently been optically 
classified by us and found to be a QSO at $z=3.12$ (Masetti et al. 2012).

Although the association between IGR J12319--0749 and this radio/X-ray high redshift QSO is intriguing, it 
cannot be considered secure as 1) there are other potential candidates in the IBIS error circle, for 
example at radio frequencies, which could emerge at higher X-ray energies and 2) it is surprising to 
detect such a distant object in a relatively short IBIS exposure (1256 ks).

For this reason, we have requested and obtained follow-up observations of IGR J12319--0749 with the 
X-ray telescope (Burrows et al. 2005) on board the \emph{Swift} satellite (Gehrels et al.2004).

\section{\emph{Swift} observations and data reduction}
During the period March 27 and July 14, 2011 XRT carried out seven observations of the 
sky region containing IGR J12319--0749 for a total exposure of $\sim$4 ks (see Table 1 for details on 
individual measurements). XRT data reduction was performed using the XRTDAS standard data pipeline 
package ({\sc xrtpipeline} v. 0.12.6), in order to produce screened event files. All data were extracted 
only in the Photon Counting (PC) mode (Hill et al. 2004), adopting the standard grade filtering (0--12 
for PC) according to the XRT nomenclature.

Events for spectral analysis were extracted within a circular region of radius 20$^{\prime \prime}$
(which encloses about 90\% of the PSF at 1.5 keV, Moretti et al. 2004) centered on the source position.
The background was extracted from various source-free regions close to the
X-ray source  using both circular/annular regions with
different radii, in order to ensure an evenly sampled background. In all
cases, the spectra were extracted from the corresponding event files using
{\sc XSELECT} software and binned using {\sc grppha} in an appropriate
way, so that the $\chi^{2}$ statistic could be applied. We used the
latest version (v.011) of the response matrices and create individual
ancillary response files (ARF) using {\sc xrtmkarf v.0.5.8}.
In all our fitting procedures we have used a Galactic
column density which in the direction of IGR J12319--0749 is 1.7$\times$ 10$^{20}$ cm$^{-2}$
(Kalberla et al. 2005).

The XRT images extracted in the 0.3--10 keV band were searched for significant excesses (above 2.5 sigma 
level) falling within the IBIS 90\% confidence error circle; only one source was clearly detected in each 
individual observation, as well as in the combined image (see Figure~\ref{fig1}). This source is located 
at R.A.(J2000) = $12^{\rm h}31^{\rm m}57^{\rm s}.67$ and Dec.(J2000) = $-07^{\circ}47^{\prime} 
19^{\prime\prime}.2$ with a positional uncertainty of $3.8^{\prime\prime}$ (90\% confidence level) which 
makes this X-ray source compatible with the faint detection reported by ROSAT. It also coincides with the 
NVSS/QSO object mentioned above; since this is the only source detected in X-rays, we assume that it is 
the true counterpart of the IBIS object.

Table~\ref{tab1} reports for each X-ray measurement, the observation date, the relative exposure, the net 
count rate in the 0.3--10 keV energy band, and the detection significance. There is some evidence that 
the source flux underwent some changes over the XRT monitoring with the most evident variation occurring 
between observation \#2 and \#7, i.e. over a few months period. Variability in QSO is not unexpected and 
can be used to characterise further this new gamma-ray source.

The UVOT instrument (Roming et al. 2005) observed IGR J12319--0749 in conjunction with the XRT for all of 
the observations listed in Table~\ref{tab1}. Each observation consists of one or more exposures with a 
single UVOT filter. The data were analysed using the latest HEASoft tools and calibration products 
following the methods described by Poole et al. (2008) and Breeveld et al. (2010). A 3$^{\prime \prime}$ 
source aperture centered on the X-ray position\footnote{When detected, the UVOT source is only 0.4 arcsec 
from the X-ray position.} and a surrounding annulus background region were used for all the exposures; 
the usual corrections including dead time, coincidence loss, and aperture loss were made. Magnitudes were 
obtained from the observed count rates using the latest zero points (Breeveld et al. 2011) and are 
reported in Table~\ref{tab1} for each observation. To convert to AB magnitudes, 1.02, 1.51, and 1.69 
should be added to the reported u, uvw1, and uvw2 magnitudes respectively (Breeveld et al. 2011). The 
source was only detected in the u filter in observation \#4 and \#7 while $90\%$ confidence upper limits 
for either the uvw1 or uvw2 filters are listed for the other measurements. Count rates were also 
converted to fluxes using the prescription of Poole et al. (2008).

\section{X-ray spectral analysis and results}
The spectral analysis of the XRT data indicated that the soft X-ray spectrum during observation \#1 and 
\#4, i.e. the two with the best statistical
significance, is hard and bright: a simple power law fit to these two measurements provides
$\Gamma=0.96^{+0.32}_{-0.34}$ and $\Gamma=1.53\pm0.28$, together with a 2--10 keV flux of
$\sim 5.7 \times 10^{-12}$ and $\sim 2.6 \times 10^{-12}$ erg cm$^{-2}$ s$^{-1}$, respectively. 
Note that for simplicity, here and in the following all modelling has been carried out in the 
observer's rest frame.

The 0.1--2.4 keV band flux is at around $1.4-1.5 \times 10^{-12}$ erg cm$^{-2}$ s$^{-1}$ to be compared 
with the lower ROSAT flux which is in the range 0.4-1 $ \times 10^{-12}$ erg cm$^{-2}$ s$^{-1}$ depending 
on the source spectrum\footnote{see http://www.xray.mpe.mpg.de/rosat/survey/rass-fsc/ for further 
information}; this comparison provides another indication that the source might be variable in soft 
X-rays.

\begin{table*}
%\begin{center}
\centering
\footnotesize
\caption{\emph{Swift} detections.}
\label{tab1}
\begin{tabular}{lccccc}
\hline
\hline
 Observation    &     Date      &  XRT Exposure$^{a}$   &   XRT Count Rate$^{b}$             &  
XRT $\sigma$  & UVOT mag \\
                &               &     (sec)         &   (10$^{-3}$ counts s$^{-1}$)  &        &      \\
\hline
\hline
\#1             &  27/03/2011   &   972             &  83.2$\pm$ 9.3                 &  8.9  &  w1$\le$21.18\\
\#2             &  10/05/2011   &   287             & 109.0$\pm$19.8                 &  5.5  &  w1$\le$21.28\\
\#3             &  25/06/2011   &   344             &  51.2$\pm$12.4                 &  4.1  &  w2$\le$21.43\\
\#4             &  28/06/2011   &  1702             &  77.7$\pm$6.8                  & 11.4  &  u=19.35$\pm$0.08\\
\#5             &  29/06/2011   &   189             &  74.0$\pm$19.8                 &  3.7  &  w2$\le$ 22.49\\
\#6             &  05/07/2011   &   129             &  51.2$\pm$20.6                 &  2.5  &  w1$\le$20.33\\
\#7             &  14/07/2011    &   341             &  37.0$\pm$10.6                &  3.5  &  u=19.81$^{+0.35}_{-0.27}$ \\
\hline
\hline
\end{tabular}
\begin{list}{}{}
\item $^{a}$ Total on-source exposure time;
\item $^{b}$ The count rate is estimated in the 0.3--10 keV energy range.
\end{list}
%\end{center} 
\end{table*}

To better characterise the QSO's X-ray spectrum and in view of the fact that the IBIS detection is
over a few revolutions, i.e provides an average flux, we have also combined all seven XRT 
spectra and repeated the spectral analysis. 
The fit to the average XRT spectrum provides a photon index $\Gamma=1.32\pm0.14$ and 
a  2--10 keV flux of $\sim$$3.3 \times 10^{-12}$ erg cm$^{-2}$ s$^{-1}$.

\begin{figure} 
\centering
\includegraphics[width=0.65\linewidth,angle=-90]{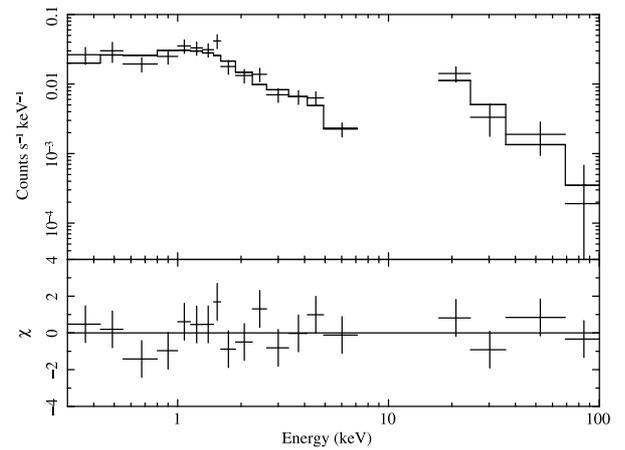}
\caption{X-ray spectrum of IGR J12319--0749 fitted with a cut-off power law.} 
\label{fig2}
\end{figure}

Finally, a joint spectral fit to the average XRT and IBIS data was also attempted. To account for possible 
calibration mismatch between the two instruments and/or variability between the two datasets, we 
introduced a constant $C$ which was left free to vary. A simple power law provides an acceptable fit with 
$\Gamma =1.35 \pm 0.14$, $C=0.8^{+0.67}_{-0.38}$ and $\chi^{2}=31.4$ for 22 d.o.f. Since in the data to 
model ratio, there is some evidence for a high energy cut-off ($E_{\rm C}$), we have also fitted the 
source broad-band spectrum with a cut-off power law: in this case, $\Gamma =1.24 ^{+0.17}_{-0.19}$, 
$C=2.53^{+5.24}_{-1.55}$, $E_{\rm C}$=$24.5^{70.6}_{-14.6}$, and the $\chi^{2} =25.7$ for 21 d.o.f. (see 
Figure~\ref{fig2}). With respect to a simple power law the $\Delta \chi^{2}$ is 5.7 for 1 extra parameter, 
which implies a fit improvement at a confidence level of around 96$\%$, not highly significant, but still 
suggestive that a spectral change might be present in the spectrum of IGR J12319--0749. In the source 
rest frame this change would be located at around 100 keV.

Assuming that this broad-band fit represents the average state of the source, we obtain observer frame 
luminosities\footnote{We adopt $H_{\rm o}=71$ km s$^{-1}$ Mpc$^{-1}$, $\Omega_{\Lambda}=0.73$ and 
$\Omega_{\rm M}=0.27$.} of $2.8\times 10^{47}$ erg s$^{-1}$ in the X-ray (2--10 keV) band and 7.3 $\times 
10^{47}$ erg s$^{-1}$ at hard X-rays (20--100 keV), i.e. IGR J12319--0749 has a high energy luminosity 
similar to those observed in the Ghisellini et al. (2010) sample of high redshifts blazars.

\section{Is IGR J12319-0749 another powerful blazar?}
The source is reported in the NED database\footnote{available at http://ned.ipac.caltech.edu/.}
 as NVSS J123157--074717, but with no 
reference, while it is 
not contained in the SIMBAD
archive; despite being potentially quite a powerful AGN, it has never been studied
before. Nonetheless, the sparse information available is sufficient to characterise broadly this new
gamma-ray source.

The NVSS (NRAO VLA Sky Survey, Condon et al. 1998) image of IGR J12319--0749 shows the source to be core 
dominated with no extended radio features and a 1.4 GHz flux of 59.8$\pm$1.8 mJy. The source is 
also listed in the FIRST survey with a similar 1.4 GHz flux (60.99$\pm$0.16 mJy, White et al. 1997). 
No other detection is reported at radio frequencies preventing a measurement of the source spectrum and an 
estimate of the radio loudness defined as $RL =L_{\rm 5GHz}/L_{\rm B}$. Despite this, we can make some 
guesses on the source spectral shape at radio frequencies considering that IGR J12319--0749 has not been 
detected in the Australian Telescope 20GHz (AT20G) survey posing a limit of 40 mJy to its flux at this 
frequency (Murphy et al. 2010). Assuming $S_{\nu} \propto \nu^{\alpha}$ and the information available, we 
estimate that $\alpha$ is $\le$ --0.15; this makes the source compatible with being a flat spectrum radio 
QSO\footnote{Flat spectrum radio sources have $\alpha$ $\ge$ --0.5 (Massardi et al. 2011)}. Furthermore, 
by evaluating the 
source B magnitude flux we should also be able to put a limit on the source radio loudness. IGR 
J12319--0749 is listed in the USNO--B1.0 catalogue (Monet et al. 2003) with a $B$ magnitude in the range 
19.7--19.8 while the $R$ and $I$ magnitudes are $\sim$19.7 and $\sim$18.2 respectively. Using the 
standard photometric system conversion from magnitude to flux (Zombeck 1990), we estimate a $B$ flux in 
the range 0.05--0.06 mJy; extrapolating the 1.4 GHz flux to 5 GHz using the limit obtained on the radio 
spectral index, it is possible to estimate loosely that the radio loudness is $\le$1000, sufficiently 
high to suggest that the source maybe a radio loud AGN\footnote{Radio loud AGN have RL values in 
the range 10--100 (Kellerman et al. 1989)}.

 IGR J12319--0749 is not reported in the near infrared 2MASS survey, with upper limits to the $J$, $H$, 
$K$ flux of $\sim$17.1, $\sim$16.4, and $\sim$15.3 magnitudes, respectively (Skrutskie et al. 2006) .

\begin{figure*} 
\includegraphics[width=8cm,height=9cm]{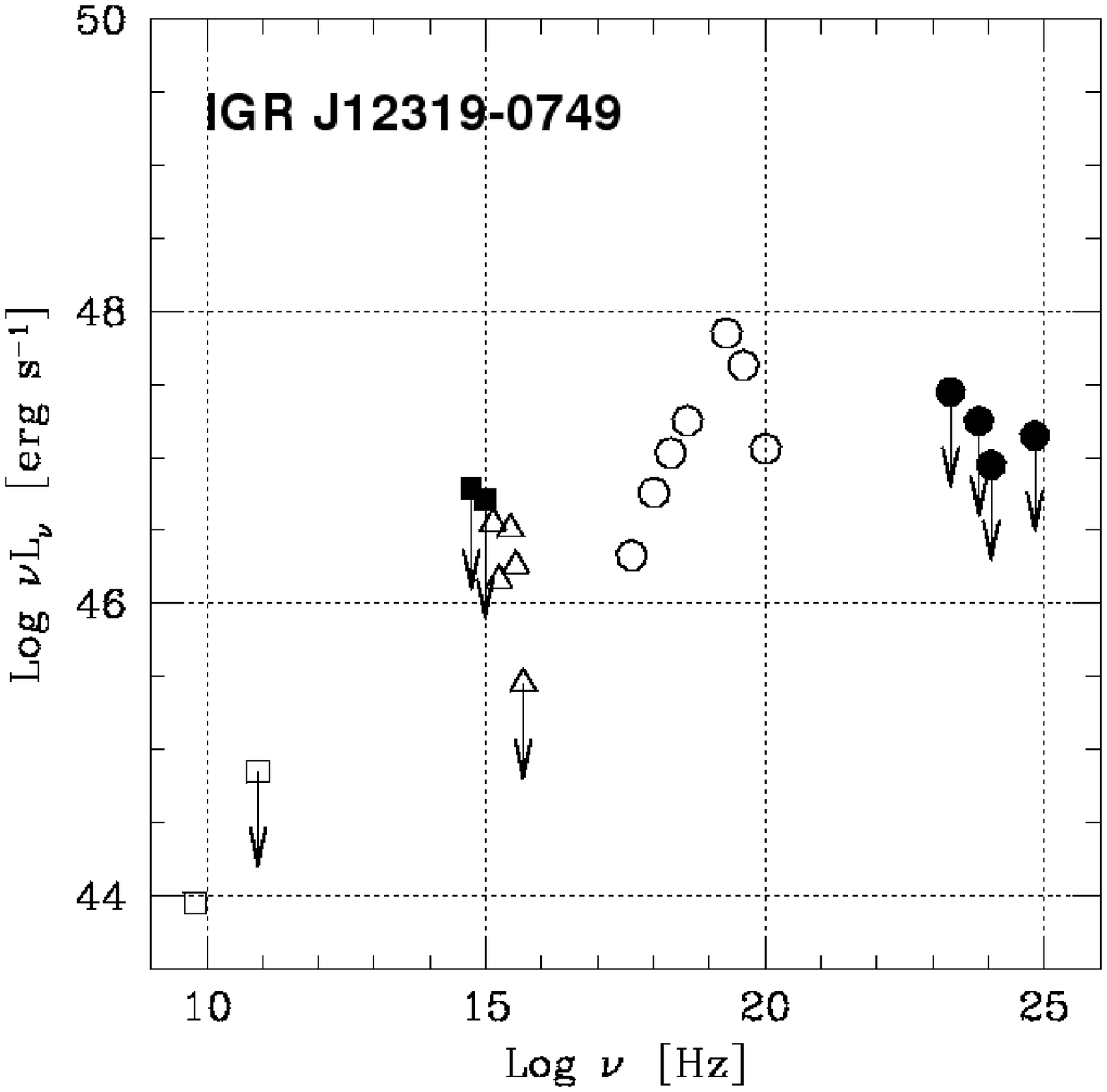}
\includegraphics[width=10cm,height=9.5cm]{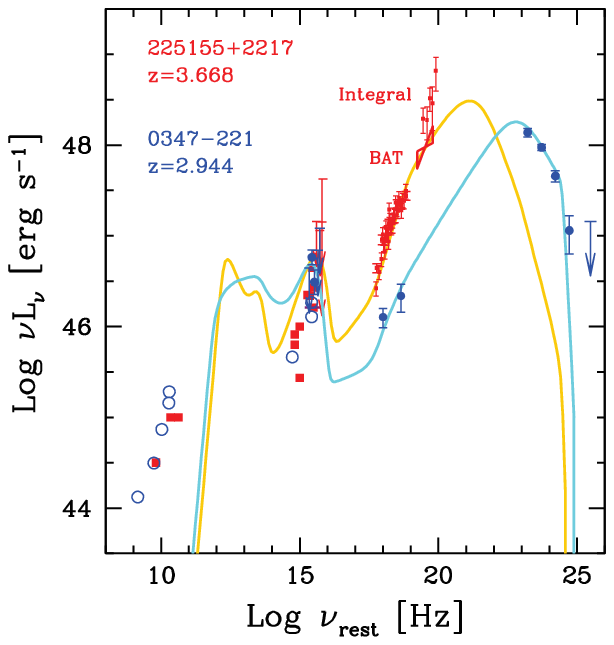}
\caption{\emph{Left}: Non simultaneous SED (source rest frame) of IGR J12319--0749; data and 
references are discussed in the text. \emph{Right}: SED (source rest frame) of two high-z QSOs 
(225155+2217 at $z=3.668$, and 0347--221 at $z=2.944$) from Ghisellini (2011).}
\label{fig3}
\end{figure*}

Optically, the source is classified as a broad emission line AGN (FWHM = 5600 km s$^{-1}$) by Masetti et 
al. (2012); these authors further estimate the mass of the black hole at the center of the QSO to be 
$2.8\times10^{9}$ solar masses, i.e. quite a massive black hole similar to those found in other 
high redshift 
blazars discovered in hard X-ray surveys (Ghisellini et al. 2010, De Rosa et al. 2012) .

At high energies, the source is bright with a hard X-ray spectral shape ($\Gamma=1.2-1.3$) and some 
indication of flux variability. There is marginal evidence in the combined XRT/IBIS data 
for a spectral break at around 100 keV (source rest frame) which could be interpreted as a peak in the 
spectral energy distribution.

Taken all together, the observed properties suggest that IGR J12319--0749 could be a bright blazar, 
in 
which the emission is relativistically beamed and the SED is double peaked. In Figure~\ref{fig3} (left 
side), we construct the non-simultaneous (source rest frame) SED of this object by combining all data 
gathered in this work; the optical UVOT data are not corrected for Galactic reddening which is however 
small in the source direction (less than 10\% in B). To cover as many frequencies as possible we have 
also used upper limits obtained from the latest \emph{Fermi} survey (Abdo et al. 2012). We adopt the 
cut-off power law model to describe the combined \emph{Swift}/XRT and \emph{INTEGRAL}/IBIS spectrum. 
Although the data are sparse and not simultaneous, the source SED resembles that of a blazar with the 
synchrotron peak located between radio and near-infrared frequencies and the Compton peak in the hard 
X-ray band. For comparison, we show on the right side of Figure~\ref{fig3} a plot taken from Ghisellini 
(2011) which describes the SED (also source rest frame) of two high-z blazars together with a single leptonic 
model fit to their data: one source (225155+2217 at $z=3.668$) has been detected by \emph{INTEGRAL}/IBIS 
and \emph{Swift}/BAT but not by \emph{Fermi}/LAT, the other (0347--221 at $z=2.944$) has been seen by LAT 
but not by IBIS or BAT. IGR J12319--0749 is more similar to 225155+2217 (see also Lanzuisi et al. 
2012) than to 0347--221. It shows a 
similar overall brightness, Compton peak location and is not detected by \emph{Fermi}/LAT.

From the available optical/UV data and reported black hole mass, it is possible to evaluate that the 
source accretion disk is emitting at 10$\%$ or more of the Eddington luminosity (see 
Figure~\ref{fig3}, right panel). 
This is similar to other high-z blazars discovered by hard X-rays surveys, which 
tend to have black holes with M greater than 10$^{9}$ solar masses and disks emitting close or above 
10$\%$ of the Eddington limit.

\section{Conclusions}
Through X-ray follow-up observations with \emph{Swift}/XRT, we have been able to likely identify 
the newly discovered \emph{INTEGRAL} source, IGR J12319--0749, with a radio source coincident with an 
AGN at $z = 3.12$. This would make IGR J12319--0749 the second most distant
object so far detected by \emph{INTEGRAL}.

The source probably belongs to the class of flat spectrum radio QSO: it is a broad line AGN with a huge 
black hole at its center; it is likely radio loud and it shows variability at X-ray energies. The source 
SED is also similar to another high-z source (225155+2217) discovered by means of hard X-ray 
observations: it has a similar brightness and disk luminosity, a Compton peak location in the hard 
X-ray soft gamma-ray domain and is not detected by \emph{Fermi}/LAT. We conclude that this is likely 
another example 
of an extreme blazar, i.e. those showing the most powerful jets, the most luminous accretion disks and 
the largest black hole masses. This finding further supports the claim that the hard X-ray waveband is 
the most efficient to discover such powerful AGN.

\begin{acknowledgements} 
We acknowledge finacial support from ASI under contract ASI I/033/10/0. This research has made use of the
NED NASA/IPAC Extragalactic Database (NED) operated by JPL (Caltech)aboratory and of the HEASARC archive
provided by NASA's Goddard Space Flight Center.
\end{acknowledgements}


\begin{thebibliography}{}

\bibitem{} Abdo, A. A., Ackermann, M., Ajello, M., et al. 2012, \apjs, in press 
(arXiv:1108.1435v1)
\bibitem{} Bird, A. J., Bazzano, A., Bassani, L., et al. 2010, \apjs, 186, 1
\bibitem{} Breeveld, A. A., Curran, P. A., Hoversten, E. A., et al. 2010, \mnras, 406, 1687 
\bibitem{} Breeveld, A. A., Landsman, W., Holland, S. T., et al. 2011, AIP Conference 
Proceedings, Vol. 1358, 373
\bibitem{} Burrows D. N.,  Hill, J. E.,  Nousek, J. A.,  et al. 2005, \ssr, 120, 165
\bibitem{} Condon, J. J., D. Cotton, W., Greisen, E. W., et al. 1998, \aj, 115, 1693
\bibitem{} De Rosa, A., Ghisellini, G., Lanzuisi, G., et al. 2012, in preparation
\bibitem{} Fossati, G., Maraschi, L., Celotti, A., Comastri, A. \& Ghisellini, G. 1998, \mnras, 
299, 433
\bibitem{} Gehrels, N., Chincarini, G., Giommi, P., et al. 2004, \apj, 611, 1005
\bibitem{} Ghisellini, G., Celotti, A., Fossati, G., et al. 1998, \mnras, 301, 451
\bibitem{} Ghisellini, G., Della Ceca, R.,  Volonteri, et al. 2010, \mnras, 405, 387
\bibitem{} Ghisellini, G. 2011, Proceedings of INTEGRAL Workshop "The Extreme and Variable High Energy Sky"(arXiv:1112.3349)
\bibitem{} Hill J. E., Burrows, D. N., Nousek, J. A., et al. 2004, Proc. SPIE, 5165, 217
\bibitem{} Kalberla, P. M. W., Burton, W. B., Hartmann, D., et al. 2005, \aap, 440, 775
\bibitem{} Kellermann, K. I., Sramek, R., Schmidt, M., Shaffer, D. B. \& Green, R. 1989, \aj, 98, 1195
\bibitem{} Lanzuisi, G., de Rosa, A., Ghisellini, G., et al. 2012, \mnras, 421, 390
\bibitem{} Malizia, A., Bassani, L., Panessa, F. 2012, in preparation
\bibitem{} Masetti, N., Parisi, P., Jimenez-Bailon, E., et al. 2012, \aap, 538, A123
\bibitem{} Massardi, M., Ekers, R. D., Murphy, T., et al. 2011, \mnras, 412, 318
\bibitem{} Monet, D. G., Levine, S. E., Canzian, B., et al. 2003, \aj, 125, 984
\bibitem{} Moretti, A., Campana, S. Tagliaferri, G., et al. 2004, Proc. SPIE, 5165, 232
\bibitem{} Murphy, T., Sadler, E.M., Ekars, R. D., et al. 2010, \mnras, 402, 2403
\bibitem{} Poole, T. S., Breeveld, A. A., Page, M. J., et al. 2008, \mnras, 383, 627  
\bibitem{} Skrutskie, M. F., Cutri, R. M., Stiening, R., et al. 2006, \aj, 131, 1163
\bibitem{} Roming, P. W. A., Kennedy, T. E., Mason, K. O., et al. 2005, \ssr, 120, 95
\bibitem{} Voges, W., Aschenbach, B., Boller, T., et al. 2000, IAU Circ., 7432
\bibitem{} White, R. L., Becker, R.H., Helfand, D. J., Gregg, M. D. 1997, \apj, 475, 479
\bibitem{} Zombeck, M. V. 1990, Handbook of Space Astronomy and Astrophysics





\end{thebibliography}
\end{document}